\documentclass[12pt]{article}

\usepackage[english]{babel}
\usepackage{amsfonts}
\usepackage{amscd}
\usepackage{amsmath}
\usepackage{amssymb}
\usepackage{latexsym}

\begin{document}

\begin{center}
\large The isotropy group of the matrix multiplication tensor \\
\medskip
\normalsize Vladimir~P.~Burichenko  \\
\medskip
\small Institute of Mathematics of the National Academy of Sciences of Belarus \\
 e-mail: vpburich@gmail.com 
\end{center}

\begin{abstract}\noindent

By an {\em isotropy group} of a tensor $t\in V_1
\otimes V_2\otimes V_3=\widetilde V$ we mean the
group of all invertible linear transformations
of $\widetilde V$ that leave $t$ invariant and
are compatible (in an obvious sense) with the
structure of tensor product on~$\widetilde V$. We
consider the case where $t$ is the structure tensor
of multiplication map of rectangular matrices.
The isotropy group of this tensor was studied in
1970s by de Groote, Strassen, and Brockett-Dobkin.
In the present work we enlarge, make more precise,
expose in the language of group actions on tensor
spaces, and endow with proofs the results previously
known. This is necessary for studying the algorithms
of fast matrix multiplication admitting symmetries.
The latter seems to be a promising new way for
constructing fast algorithms.

({\em MSC classification 68Q25, 20C}).
\end{abstract}
\bigskip

(Warning: this is a copy of the article published by the author in 2016, so some 
of the references may be obsolete.) 

\textbf{1. Introduction.} The present article
is related to the problem of fast matrix multiplication.
In [1], [2] the author put forward an idea that
good (i.e. fast) algorithms must have nontrvial
symmetries, and this may be a fruitful way of
searching for new algorithms. To confirm this idea,
in [1] and [2] it was shown that some well-known
algorithms (Strassen's, Hopcroft's and Laderman's
[3], [4], [5]) have rather large automorphism
groups. (The definition of automorphism group of
an algorithm was also given in [1], [2]).

The automorphism group of an algorithm is a subgroup
of a certain ambient group, namely the isotropy
group of the tensor associated with operation
of matrix multiplication. Studying the latter
group is a necessary part of studying symmetries
of algorithms. However, this is a purely algebraic
problem; it is not difficult, but its solution is
not short. This is the subject of this article.

The algorithms themselves are not considered in the
article. They will be considered in the future
article [6] (and were considered in [1], [2]).
The present article is a preparation for [6], quite
similarly to the way how [7] was a preparation
article for~[8]. Thus, the aim of the present
article is to liberate the reader of [6] of some
necessary, but standard algebra.

It should be said from the very beginning that
the present article is not especially original.
The results themselves are not neither new, nor
difficult. They were mainly known 40 years ago.
See [7, Sect.3], and [9, Sect.4]. We have an
intention to make them more precise, to enlarge
them to necessary extent, to expose them in an
appropriate language (group actions on tensor
spaces), which is necessary for further applications,
and to endow them with proofs.

J.M.Landsberg communicated to the author that he (with co-authors)
independently came to the idea of using symmetry for analisys and
constructing matrix multiplication algorithms, approximately at
the same time as the author of the present article. See preprint
[10] for more details. Also, it should be mentioned that symmetry
groups were used in [15] to find some new algorithms. However, we
want to warn the reader on the following. In the listed works,
where group action on algorithms, or symmetry of algorithms, were
considered ({1],[2], [7],[9],[10],[15]), every author develops his
own system of concepts, and uses his own language (so that, it is
possible that in the present work the author lays foundations only
for his own future work).

Finally, inform the reader that this article is
a revised version of Sections 3 and 4 of~[2].
However, the differences with [2] are significant.
Some points are exposed in [2] with more details
and explanations. On the othere hand, some places
in [2] were too complicated and difficult, and in
this article the author tried to simplify them.

\textbf{2. The isotropy group of a tensor.}
We assume the
reader is familiar with the basics of multilinear
algebra and group representation theory, including
the concept of tensor product of representations.
See [11, Ch.4], [12, Ch.8], [13,Ch.1]. (This remark
is made for possible readers who are not pure
matematicians, but, for example, computer scientists).

Let $K$ be a field, $\widetilde V=V_1\otimes\ldots\otimes V_l$ a
tensor product of several spaces. By a {\em decomposable
automorphism} we mean any invertible transformation $\varphi\in
GL(\widetilde V)$, which is compatible, in an obvious sense, with
the structure of a tensor product on $\widetilde V$. For example,
let $\widetilde V=V_1\otimes V_2\otimes V_3$, and let
$\alpha:V_1\longrightarrow V_2$, $\beta:V_2\longrightarrow V_1$,
and $\gamma:V_3\longrightarrow V_3$ be some isomorphisms (so that
necessary ${\rm dim\,} V_1= {\rm dim\,} V_2$).  Then the
transformation of $\widetilde V$, defined by
$$ v_1\otimes v_2\otimes v_3\mapsto \beta(v_2)\otimes\alpha(v_1)
\otimes\gamma(v_3), $$
is a decomposable automorphism.

The group of all decomposable automorphisms of
$\widetilde V$ will be denoted by $S(V_1,\ldots,V_l)$.
Those that preserve all factors $V_i$ form a
normal subgroup, denoted by $S^0(V_1,\ldots,V_l)$.
In other words, $S^0(V_1,\ldots,V_l)$ is the group
of all transformations of the form $g_1\otimes
\ldots\otimes g_l$, where $g_i\in GL(V_i)$.
The following statement holds.

\textbf{Proposition 1.} {\it
The group $S^0=S^0(V_1,\ldots,V_l)$
is a central product of the groups $GL(V_i)$,
$i=1,\ldots,l$. More precisely, $S^0\cong A/B$,
where $A=GL(V_1)\times\ldots\times GL(V_l)$
and $B$ is the subgroup of all elements of the
form $(\lambda_1{\rm id}_{V_1},\ldots,\lambda_l
{\rm id}_{V_l})$, where $\lambda_i\in K^\ast$ and
$\lambda_1\ldots\lambda_l=1$. The quotient group
$S(V_1,\ldots,V_l)/S^0$ may be naturally identified
with the group of all permutations of the set
$\{V_i\mid {\rm dim\,} V_i>1\}$, preserving dimensions. }

The proof of this proposition is not difficult, but
a bit tedious. It is left to the reader, or can be
found in [2, Sect.3].

\textbf{Definition.} Let $\widetilde V=V_1\otimes\ldots\otimes V_l$, and
let $t\in\widetilde V$ be an arbitrary non-zero tensor.
The groups
$$ \Gamma(t)=\{g\in S(V_1,\ldots,V_l)\mid g(t)=t\}$$
and
$$ \Gamma^0(t)=\Gamma(t)\cap S^0(V_1,\ldots,V_l)$$
are called {\em the isotropy group} and {\em the
small isotropy group} of $t$, respectively.

Let $M_{a,b}(K)$ be the space of all $a\times b$
matrices over~$K$. Its usual basis is
$\{e_{ij}\mid 1\leq i\leq a,\, 1\leq j\leq b\}$,
where $e_{ij}$ are the usual matrix units. We will
briefly denote $M_{a,b}(K)$ by $M_{ab}$, and the
group $GL_n(K)$ by $GL_n$. Also we will use the
notation $\overline n=\{1,\ldots,n\}$.

Let $m,n,p\in{\Bbb N}$. Consider the product
$L=M_{mn}\otimes M_{np}\otimes M_{pm}$. In the theory of
fast matrix multiplication a very important role
is played by the following tensor, often denoted
by $\langle m,n,p\rangle$:
$$ \langle m,n,p\rangle=\sum_{i,j,k} e_{ij}\otimes e_{jk}
\otimes e_{ki}\,\in L, $$
where the sum is over all $i\in\overline m$, $j\in\overline n$,
$k\in\overline p$.

To describe the group $\Gamma(\langle m,n,p\rangle)$
is the main goal of the present work.

\textbf{3. A subgroup of $\Gamma^0$.}  Let $C_l$ and $R_l$
denote the spaces of all columns of height $l$,
resp. the rows of length $l$, over $K$, and let
$(e_i\mid i\in\overline l)$ and $(e^i\mid i\in\overline l)$
be the usual bases in $C_l$ and $R_l$, respectively.
If $c\in C_l$ and $r\in R_l$, then $rc$ is a
$1\times1$ matrix, i.e. a scalar. The map $(r,c)
\mapsto rc$ is a pairing (that is, a nondegenerate
bilinear map), and $(e_i)$ and $(e^i)$ are dual
bases. Thus, we can consider $C_l$ and $R_l$ as
dual spaces.

The group $G=GL_l$ acts on $C_l$ as usually:
$(g,v)\mapsto gv$, where $gv$ is the usual product
of a matrix by a column. Also, there is a left
action of $G$ on $R_l$ by
$$ (g,v')\mapsto g\circ v':=v'g^{-1} \,.$$
(This is a left action indeed, that is, $(gh)
\circ v'=g\circ(h\circ v')$ for all $g,h\in G$
and $v'\in V'$. Indeed, $g\circ(h\circ v')=
g\circ(v'h^{-1})=(v'h^{-1})g^{-1}=v'h^{-1}
g^{-1}=v'(gh)^{-1}=(gh)\circ v'$.). So there is
a left action of $G$ on $C_l\otimes R_l$ such that
$$ g(v\otimes v')=gv\otimes v'g^{-1}\,,\qquad
\forall g\in G,
\ v\in C_l,\  v'\in R_l. $$

Consider the tensor
$$ \delta=\delta_{(l)}=\sum_{i=1}^le_i\otimes e^i
\in C_l\otimes R_l $$
(so-called identity tensor). The next lemma is
standard; nevertheless we give a proof.

\textbf{Lemma 2.} {\it
We have $g\delta=\delta$, for all $g\in G$. }

\textbf{Proof.} Let $a_{ij}$ and $b_{ij}$ be the coefficients of the
matrices $g$ and $g^{-1}$, i.e.,
$$ g=\sum_{i,j=1}^l a_{ij}e_{ij}\quad {\rm and}\quad
g^{-1}=\sum_{i,j=1}^l b_{ij}e_{ij}.$$
Then $ge_i=\sum_{j=1}^l a_{ji}e_j$ and $e^ig^{-1}=
\sum_{j=1}^l b_{ij}e^j$. Hence
\begin{eqnarray*}
g\delta &=& g(\sum_{i=1}^le_i\otimes e^i)= \sum_{i=1}^l
ge_i\otimes e^ig^{-1}=\sum_{i=1}^l(\sum_{j=1}^l a_{ji}e_j)
\otimes(\sum_{k=1}^lb_{ik}e^k) \\
&=& \sum_{j,k=1}^l(\sum_{i=1}^l a_{ji}b_{ik})e_j
\otimes e^k=\sum_{j,k=1}^l(\delta_{jk})e_j\otimes e^k=
\sum_{j=1}^le_j\otimes e^j=\delta,
\end{eqnarray*}
as $\sum_{i=1}^l a_{ji}b_{ik}=\delta_{jk}$ for all $1\leq j,k
\leq l$ (because matrices $g$ and $g^{-1}$ are inverse).
\hfill $\square$ \medskip

Let $m,n,p\in{\Bbb N}$, and $L=L_1\otimes L_2\otimes L_3$, where
$L_1=M_{mn}$, $L_2=M_{np}$, $L_3=M_{pm}$. Define certain action of
$G=GL_m\times GL_n \times GL_p$ on~$L$. For $(a,b,c)\in G$ define
the transformation $T(a,b,c)$ of $L$ by the formula
$$ T(a,b,c)(x\otimes y\otimes z)=axb^{-1}\otimes byc^{-1}
\otimes cza^{-1}\,. $$
It is easy to see that the rule $g\mapsto T(g)$
is a homomorphism from $G$ to $GL(L)$, that is
always $T(a_1,b_1,c_1)T(a_2,b_2,c_2)=T(a_1a_2,
b_1b_2,c_1c_2)$, and $T(E_m,E_n,E_p)={\rm id}_L$.

Indeed, for any $x\otimes y\otimes z\in L$ we have
\begin{eqnarray*} T(a_1,b_1,c_1)(T(a_2,b_2,c_2)
(x\otimes y\otimes z)) &=& T(a_1,b_1,c_1)
(a_2xb_2^{-1}\otimes b_2yc_2^{-1} \otimes
c_2za_2^{-1})\\
&=& a_1a_2xb_2^{-1}b_1^{-1}\otimes b_1b_2y
c_2^{-1}c_1^{-1} \otimes c_1c_2za_2^{-1}a_1^{-1}\\
&=& (a_1a_2)x(b_1b_2)^{-1}\otimes (b_1b_2)y
(c_1c_2)^{-1} \otimes (c_1c_2)z(a_1a_2)^{-1}\\
&=& T(a_1a_2,b_1b_2,c_1c_2)(x\otimes y\otimes z).
\end{eqnarray*}

\textbf{Proposition 3.} {\it The transformations
$T(a,b,c)$ preserve $t=\langle m,n,p\rangle$. }

\textbf{Proof.} Since $g\mapsto T(g)$ is a homomorphism,
and $G=GL_m\times GL_n\times GL_p$ is a direct
product, it suffices to prove that $t$ is invariant
under $T(g)$ if $g$ is in one of the factors
$GL_m$, $GL_n$, or $GL_p$. For instance, let $g\in
GL_m$, that is, more precisely, $g=(a,E_n,E_p)$,
where $a\in GL_m$. Let $a'=(a'_{ij})=a^{-1}$ be
the matrix inverse to~$a$. Then
$$ T(g)t=T(a,E_n,E_p)\sum_{(i,j,k)\in\overline m
\times\overline n \times\overline p} e_{ij}
\otimes e_{jk}\otimes e_{ki}= \sum_{(i,j,k)\in
\overline m\times\overline n\times\overline p}
ae_{ij}\otimes e_{jk}\otimes e_{ki}a'\,.$$

It is sufficient to prove that for all $(j,k)\in
\overline n\times\overline p$ the sum of all summands in
$T(g)t$ having $e_{jk}$ in the middle coincides
with the similar sum in $t$, that is,
$$ \sum_{i=1}^m ae_{ij}\otimes e_{jk}\otimes e_{ki}a'
= \sum_{i=1}^m e_{ij}\otimes e_{jk}\otimes e_{ki}\,. $$
We have
$$ a=\sum_{r,s=1}^m a_{rs}e_{rs}\,,\qquad
a'=\sum_{r,s=1}^m a'_{rs}e_{rs}\,, $$
whence
$$ ae_{ij}=\sum_{r=1}^m a_{ri}e_{rj}\,,\qquad
e_{ki}a'=\sum_{s=1}^ma'_{is}e_{ks}\,.$$
So
\begin{eqnarray*} \sum_{i=1}^m ae_{ij}\otimes
e_{jk}\otimes e_{ki}a' &=& \sum_{i,r,s=1}^m a_{ri}
a'_{is} e_{rj}\otimes e_{jk}\otimes e_{ks}=
\sum_{r,s=1}^m\delta_{rs} e_{rj}\otimes
e_{jk}\otimes e_{ks} \\
&=& \sum_{r=1}^m e_{rj}\otimes e_{jk}
\otimes e_{kr}= \sum_{i=1}^m e_{ij}\otimes e_{jk}
\otimes e_{ki}\,,
\end{eqnarray*}
as required. Here we have used the equality
$\sum_{i=1}^m a_{ri}a'_{is}=\delta_{rs}$, as $a$
and $a'$ are inverse matrices.
\hfill $\square$ \medskip

(There is a less computational and more conceptual
proof, which is, in brief, as follows.

If $c\in C_r$ and $r\in R_s$, then $cr\in M_{rs}$ is an $r\times
s$ matrix. Now we consider the product
$$ N=C_m\otimes R_m\otimes C_n\otimes R_n\otimes C_p\otimes R_p $$
and the linear map $\tau:N\longrightarrow L$
defined by
$$ c_1\otimes r_1\otimes c_2\otimes r_2\otimes c_3\otimes r_3\mapsto
c_1r_2\otimes c_2r_3\otimes c_3r_1 $$
(as the expression $c_1r_2\otimes c_2r_3\otimes c_3r_1$
is linear in each of the arguments $c_1,\ldots,r_3$,
this is a well-defined map indeed). It is easy to see that
this is an isomorphism of vector spaces.

We can consider the ``identity tensors''
$\delta_{(m)}\in C_m\otimes R_m$, $\delta_{(n)}$, and
$\delta_{(p)}$. Then $\delta_{(m)}\otimes\delta_{(n)}
\otimes\delta_{(p)}$ is an element of~$N$. It is easy
to see that $\tau(\delta_{(m)}\otimes\delta_{(n)}
\otimes\delta_{(p)})$ is nothing else but $\langle m,n,p
\rangle$.

There is an action of $G=GL_m\times GL_n\times
GL_p$ on $N$ by the rule
$$(a,b,c)(c_1\otimes r_1\otimes c_2\otimes r_2\otimes c_3\otimes r_3)
=ac_1\otimes r_1a^{-1}\otimes bc_2\otimes r_2b^{-1}\otimes cc_3
\otimes r_3c^{-1}.$$
It is possible to check that $\tau$ is a
$G$-homomorphism, with respect to this action.
Next, it can be deduced from Lemma~2 that
$\delta_{(m)}\otimes\delta_{(n)}\otimes\delta_{(p)}$
is an invariant element. So its image
$\langle m,n,p\rangle$ is an invariant element too. )
\medskip

Thus, we have the following statement.

\textbf{Proposition 4.} {\it The group $H$ of all
transformations of the form $T(a,b,c)$ is a
subgroup of $\Gamma^0(t)$. }

The inverse inclusion is also true, but to prove
it is more difficult. We prove it later in this
article.

\textbf{Proposition 5.} {\it Any element of
$\Gamma^0(t)$ has
the form $T(a,b,c)$, for some $(a,b,c)\in GL_m
\times GL_n\times GL_p$, and therefore
$\Gamma^0(t)=H$. }

The representation of an element of $H$ in the
form $T(g)$, $g=(a,b,c)$, is not unique. Evidently,
$T(g_1)=T(g_2)$ if and only if $T(g_2g_1^{-1})
=1$ ($={\rm id}_L$; we will sometimes write $1$
for identity map, or the identity element
of a group).

Let us find out when $T(h)=1$. We need two auxiliary
statements, whose proofs are left to the reader.

(1) {\it If $\widetilde V=V_1\otimes\ldots\otimes V_l$ is any tensor
product, then two nonzero decomposable tensors
$u_1\otimes\ldots\otimes u_l$ and $v_1\otimes
\ldots\otimes v_l$ coincide if and only if
$v_i=\lambda_iu_i$, $\lambda_i\in K^\ast$,
$\lambda_1\ldots\lambda_l=1$.}

(2) {\it Let $a\in GL_m$, $b\in GL_n$, and suppose that
$axb$ is proportional to $x$ for all $x\in M_{mn}$.
Then both $a$ and $b$ are scalar matrices. }

\textbf{Proposition 6.} {\it $T(h)=1$ if and only if $h=(\lambda
E_m, \mu E_n\,,\nu E_p)$, where $\lambda,\mu,\nu
\in K^\ast$. }

\textbf{Proof.} If $h$ is of this form, then
$$ T(h)(x\otimes y\otimes z)=\lambda x\mu^{-1}\otimes \mu y
\nu^{-1}\otimes \nu z\lambda^{-1}=x\otimes y\otimes z\,,$$
for all $x$, $y$, and $z$, whence $T(h)=1$.

Conversely, let $h=(a,b,c)$ and $T(h)=1$. Then
$axb^{-1}\otimes byc^{-1}\otimes cza^{-1}=x\otimes y\otimes z$,
for all $x\in L_1$, $y\in L_2$, $z\in L_3$. It
follows from (1) that $axb^{-1}\sim x$, for all~$x$.
Now (2) implies that both $a$ and $b$ are scalar
matrices. Similarly $c$ is a scalar matrix also.
\hfill $\square$ \medskip

\textbf{Corollary 7.} {\it $T(a_1,b_1,c_1)=T(a_2,b_2,c_2)$ if and
only if $a_2=\lambda a_1$, $b_2=\mu b_1$, $c_2 =\nu c_1$, for some
$\lambda,\mu,\nu\in K^\ast$. }

Now we can decribe the structure of $\Gamma^0(t)$
as an abstract group. Recall that the {\em
projective general linear group} is $PGL_n(K)=
GL_n(K)/Z_n(K)$, where $Z_n(K)=\{\lambda E_n\mid
\lambda\in K^\ast\}$ is the group of all scalar
matrices of size~$n$.

\textbf{Proposition 8.} {\it $\Gamma^0(t)\cong PGL_m(K)\times
PGL_n(K)\times PGL_p(K)$. }

\textbf{Proof.} We have $\Gamma^0(t)=H$ by
Proposition~5.
The map $g\mapsto T(g)$ is a surjective
homomorphism of the group $G=GL_m\times GL_n\times
GL_p$ onto~$H$. So $H$ is isomorphic to the
quotient group $H/B$, where $B=\{h\mid T(h)=1\}$.
It follows from Proposition 6 that $B=Z_m(K)\times
Z_n(K)\times Z_p(K)$. Finally, it is easy to see
that the quotient group of $G$ by the latter
subgroup is isomorphic to $PGL_m(K)\times
PGL_n(K)\times PGL_p(K)$.
\hfill $\square$ \medskip

\textbf{4. Structure of $\Gamma(t)$.}
The full isotropy group
$\Gamma(t)$, where $t=\langle m,n,p\rangle$, may be larger
than $\Gamma^0(t)$. However, the relations
between $\Gamma(t)$ and $\Gamma^0(t)$ can be easily
described.

In this section we assume that at most one of the
three numbers $m$, $n$, and $p$ is equal to $1$.
Then $mn,np, pm>1$.

First assume that $m$, $n$, and $p$ are pairwise
distinct. Then ${\rm dim\,} L_1=mn$, ${\rm dim\,} L_2=np$, and
${\rm dim\,} L_3=mp$ are pairwise distinct also. So
$S(L_1,L_2,L_3)=S^0(L_1,L_2,L_3)$, whence
$\Gamma(t)=\Gamma^0(t)$.

Next assume that $|\{m,n,p\}|=2$. We consider the
case $m=n\ne p$ only; the remaining two cases
$m=p\ne n$ and $m\ne n=p$ can be obtained from
this case in an obvious way. Define $\rho_{(23)}:
L\longrightarrow L$ by
$$ \rho_{(23)}(x\otimes y\otimes z)=x^t\otimes z^t\otimes y^t$$
(we use the same symbol $t$ for the tensor $t=
\langle m,n,p\rangle$ and the transpose map, but hope
this will not lead to a confusion). Note that
$\rho_{(23)}$ is well-defined, because operation of
taking the transpose matrix maps the spaces
$L_2=M_{np}=M_{mp}$ and
$L_3=M_{pm}$ each onto the other, and $L_1=M_{mm}$
onto itself. Observe next that $\rho_{(23)}^2=1$
($={\rm id}_L$), as
\begin{eqnarray*} \rho^2_{(23)}(x\otimes y\otimes
z) &=& \rho_{(23)} (\rho_{(23)}(x\otimes y\otimes
z))=\rho_{(23)}(x^t\otimes z^t\otimes y^t) \\
&=& (x^t)^t\otimes (y^t)^t\otimes (z^t)^t =x
\otimes y\otimes z.
\end{eqnarray*}
Finally, we have $\rho_{(23)}\in\Gamma(t)$,
because
$$\rho_{(23)}(t)=\rho_{(23)}(\sum_{\substack{
1\leq i,j\leq m \\ 1\leq k\leq p}} e_{ij}\otimes
e_{jk}\otimes e_{ki}) =\sum_{\substack{ 1\leq i,j
\leq m \\ 1\leq k\leq p}} e_{ji}\otimes e_{ik}\otimes
e_{kj}=t\,.$$

To formulate the statement on the structure of
$\Gamma(t)$, it will be convenient to use the
notion of semidirect product.
Recall that a group $G$ is the {\em product} of
its subgroups $A$ and $B$, which is denoted by
$G=AB$, if for each $g\in G$ there exist $a\in A$
and $b\in B$ such that $g=ab$. If in addition
$A\cap B=1$, then it is easy to see that the
representation of $g$ in the form $g=ab$ is unique.
Finally, a group $G$ is said to be a
{\em semidirect
product} of $A$ by $B$, which is denoted by
$G=A\leftthreetimes B$, if $G=AB$, $A$ is normal in $G$,
and $A\cap B=1$.

Let $Q=\{1,\rho_{(23)}\}$ be the subgroup of
$\Gamma(t)$ of order $2$ generated by~$\rho_{(23)}$.
Show that $\Gamma(t)=\Gamma^0(t)\leftthreetimes Q$.
We have $\Gamma^0(t)\trianglelefteq\Gamma(t)$ from
the definition of $\Gamma^0(t)$, because $S^0(V_1,\ldots,
V_l)\trianglelefteq S(V_1,\ldots, V_l)$.
Next, $Q\cap\Gamma^0(t)=1$, because $\rho_{(23)}$ induces a
nontrivial permutation of factors. Finally
$\Gamma(t)=\Gamma^0(t)Q$. Indeed, let $x\in\Gamma(t)$.
Then the permutation $\pi$, induced by $x$ on the
factors $\{L_1,L_2,L_3\}$, preserves the
dimensions, whence $\pi=1$ or $\pi=(23)$. If
$\pi=1$, then $x\in\Gamma^0(t)$. If $\pi=(23)$,
then the element
$x'=x\rho_{23}$ is in $\Gamma(t)$ and induces the
trivial permutation, whence $x'\in\Gamma^0(t)$
and $x=x'\rho_{(23)}\in\Gamma^0(t)Q$.

It remains to consider the case $m=n=p$. Define
$$ \rho_{(12)}:x\otimes y\otimes z\mapsto y^t\otimes x^t\otimes z^t.$$
Then $\rho_{(12)}^2=1$ and $\rho_{(12)}\in\Gamma(t)$
similarly to $\rho_{(23)}$. Consider the group
$Q=\langle\rho_{(12)},\rho_{(23)}\rangle$. A direct checking,
left to the reader, shows that $Q\cong S_3$ and
that any permutation of the factors $L_1$, $L_2$,
and $L_3$ is induced by a unique element of~$Q$.
It follows, quite similarly to the case $m=n\ne p$,
that $\Gamma(t)=\Gamma^0(t)\leftthreetimes Q$.

It may be useful to have explicit formulae for
conjugation of an element of $H$ by an element
of~$Q$. For a matrix $x\in GL_l(K)$ we denote
by $x^\vee$ the matrix $x^\vee=(x^t)^{-1}=
(x^{-1})^t$ (which is usually called the matrix
{\em cotragradient} to~$x$).

\textbf{Proposition 9.} {\it If $m=n$, then
$$ \rho_{(23)}T(a,b,c)\rho_{(23)}=T(b^\vee,
a^\vee,c^\vee). $$

If $m=n=p$, and for a permutation $\pi\in S_3$
$\rho_\pi$ is the element of $Q=\langle \rho_{(23)},
\rho_{(12)}\rangle$ inducing this permutation on
$\{L_1,L_2,L_3\}$, then in addition the following
relations hold:
$$ \rho_{(12)}T(a,b,c)\rho_{(12)} =T(c^\vee,
b^\vee,a^\vee), $$
$$ \rho_{(13)}T(a,b,c)\rho_{(13)}=T(a^\vee,
c^\vee,b^\vee), $$
$$ \rho_{(123)}T(a,b,c)\rho_{(123)}^{-1}=
T(c,a,b), $$
$$ \rho_{(132)}T(a,b,c)\rho_{(132)}^{-1}=
T(b,c,a). $$ }

\textbf{Proof.}  We prove the relation for $\rho_{(12)}$
as an example. Note that $\rho_{(12)}^{-1}=
\rho_{(12)}$, as $\rho_{(12)}^2=1$.
For $x\in L_1$, $y\in L_2$, and $z\in L_3$ we have
$\rho_{(12)}(x\otimes y\otimes z)=y^t\otimes x^t
\otimes z^t$, whence
\begin{eqnarray*}
x\otimes y\otimes z  && \stackrel{\rho_{(12)}}\mapsto y^t\otimes
x^t\otimes z^t \stackrel{T(a,b,c)}\mapsto ay^tb^{-1}
\otimes bx^tc^{-1}\otimes cz^ta^{-1} \\
&& \stackrel{\rho_{(12)}}\mapsto (bx^tc^{-1})^t \otimes
(ay^tb^{-1})^t \otimes (cz^ta^{-1})^t
= (c^{-1})^txb^t\otimes(b^{-1})^tya^t\otimes (a^{-1})^tzc^t \\
&& =c^\vee x(b^\vee)^{-1} \otimes b^\vee y(a^\vee)^{-1} \otimes
a^\vee z(c^\vee)^{-1} =T(c^\vee, b^\vee,a^\vee) (x\otimes y\otimes
z),
\end{eqnarray*}
whence
$$\rho_{(12)}T(a,b,c)\rho_{(12)}=T(c^\vee, b^\vee,
a^\vee).$$
\hfill $\square$ \medskip

We summarize the statements obtained so far in
the next theorem, which is the main result of the
present work.

\textbf{Theorem.} {\it Let $m,n,p\in{\Bbb N}$,
$(m,n,p)\ne(1,1,1)$,
let $L_1=M_{mn}=M_{mn}(K)$, $L_2=M_{np}$,
$L_3=M_{pm}$, let $L=L_1\otimes L_2\otimes L_3$, and let
$$ t=\langle m,n,p\rangle= \sum_{1\leq i\leq m, \
1\leq j\leq n, \ 1\leq k\leq p} e_{ij}\otimes e_{jk}
\otimes e_{ki} \in L.$$

For elements $a\in GL_m(K)$, $b\in GL_n(K)$,
$c\in GL_p(K)$ define the transformation $T(a,b,c):
L\longrightarrow L$ by the formula
$$T(a,b,c)(x\otimes y\otimes z)= axb^{-1}\otimes byc^{-1}\otimes
cza^{-1}.$$
Put
$$ H=\{ T(a,b,c)\mid (a,b,c)\in GL_m(K)\times
GL_n(K)\times GL_p(K) \}. $$
Then $\Gamma^0(t)=H$. The transformations $T(a,b,c)$
and $T(a_1,b_1,c_1)$ are equal if and only if
$a_1=\lambda a$, $b_1=\mu b$, and $c_1=\nu c$,
for some $\lambda,\mu,\nu\in K^\ast$. The group
$H$ is isomorphic to $PGL_m(K)\times PGL_n(K)
\times PGL_p(K)$.

If $m$, $n$, and $p$ are pairwise distinct, then
$\Gamma(t)=\Gamma^0(t)$.

If $m=n$ or $m=p$, then we define the trnsformations
$\rho_{(23)},\rho_{(12)}:L\longrightarrow L$ by
$$ \rho_{(23)}(x\otimes y\otimes z)=x^t\otimes z^t\otimes y^t\,,$$
$$ \rho_{(12)}(x\otimes y\otimes z)=y^t\otimes x^t\otimes z^t\,,$$
respectively. In the case $m=n\ne p$ put
$Q=\langle \rho_{(23)}\rangle$, and in the case $m=n=p$
put $Q=\langle\rho_{(12)},\rho_{(23)}\rangle$. Then
$Q\cong Z_2$ or $Q\cong S_3$ in the first and
second case, respectively. $Q$ is a subgroup of
$\Gamma(t)$, and $\Gamma(t)=\Gamma^0(t)\leftthreetimes Q$.
Any permutation of the factors $L_1$, $L_2$, $L_3$,
preserving the dimensions, is induced by a unique
element of~$Q$.

In the case $m=n$ the relation
$$ \rho_{(23)}T(a,b,c)\rho_{(23)}=T(b^\vee, a^\vee,
c^\vee), $$
holds, and in the case $m=n=p$ the relations
$$ \rho_{(12)}T(a,b,c)\rho_{(12)} =T(c^\vee,
b^\vee,a^\vee), $$
$$ \rho_{(13)}T(a,b,c)\rho_{(13)}=T(a^\vee,
c^\vee,b^\vee), $$
$$ \rho_{(123)}T(a,b,c)\rho_{(123)}^{-1}=
T(c,a,b), $$
$$ \rho_{(132)}T(a,b,c)\rho_{(132)}^{-1}=
T(b,c,a) $$
holds also. Here $\rho_\pi$ is the element of $Q$
inducing the permutation $\pi\in S_3$ on the
factors.

In the cases $m=p\ne n$ and $n=p\ne m$ the
statements similar to those for the case $m=n\ne p$
are true.
}

\bigskip
The rest of the article is devoted to the
proof of Proposition~5.
\medskip

\textbf{5. Transformations of matrix spaces.}
There is a well-known theorem stating that any
automorphism of the algebra of square matrices
over a field is induced by the conjugation by a
nondegenerate matrix. In this section we prove
an analogue of this theorem for the multiplication
of rectangular matrices.

By $R_n$ and $C_n$ we denote the spaces of all
rows of length $n$, respectively the columns of
height $n$, over a given field $K$; i.e. $R_n(K)
=M_{1n}(K)$ and $C_n=M_{n1}(K)$.
Note that if $c\in C_m$ and $r\in R_n$, then
$cr\in M_{mn}$. Moreover, the rule $c\otimes r\mapsto
cr$ defines an isomorphism $C_m\otimes R_n\longrightarrow M_{mn}$.
Also, observe that ${\rm rk}(X)=1$ if and only if
$X=cr$ for some $C\in C_m$ and $r\in R_n$.
Here ${\rm rk}$ is the rank of a matrix. Finally
note that for any $c\in C_l$ and $r\in R_l$ the
product $rc$ is a $1\times1$ matrix, that is,
a scalar.

Recall that a (non-zero) element of a tensor
product $\widetilde V=V_1\otimes\ldots\otimes V_l$ of the form
$v_1\otimes\ldots\otimes v_l$ is called a {\em decomposable
tensor}. It is clear that if $\widetilde U=U_1\otimes\ldots
\otimes U_l$, $\widetilde V=V_1\otimes\ldots\otimes V_l$, and
$\varphi:\widetilde U\longrightarrow\widetilde V$
is a decomposable isomorphism,
then $\varphi$ takes decomposable tensors to
decomposable ones. The converse is also true:

\textbf{Proposition 10.} {\it Let $\widetilde U=U_1\otimes\ldots\otimes U_l$,
$\widetilde V=V_1\otimes\ldots\otimes V_l$, $X$ and $Y$ be the
sets of all decomposable tensors in $\widetilde U$ and
$\widetilde V$ respectively, and $\varphi:\widetilde U\longrightarrow\widetilde V$ be
an isomorphism of linear spaces that bijectively
maps $X$ onto~$Y$. Then $\varphi$ is a decomposable
isomorphism. }

A proof of this proposition is contained in~[1].
It is not difficult. Another proof can be found
in~[7]. Anyway, this statement, no doubt, is an
old result of the classical projective algebraic
geometry (``automorphisms of Segre embeddings'').
(However, the author could not find it, in an
explicit form, in the available textbooks.
In [7] the book [14] is mentioned, which is not available to
the author.) So, we will not prove this statement
here.

\textbf{Proposition 11.} {\it Let $A$ be an
invertible linear
transformation of the space $M_{mn}$, having
the property that ${\rm rk}(Ax)=1$ for all $x$ such
that ${\rm rk}(x)=1$. Then either there exist
$a\in GL_m$ and $b\in GL_n$ such that $Ax
=axb$ for all $x\in M_{mn}$, or $m=n$ and there
exist $a,b\in GL_n$ such that $Ax=ax^tb$
for all $x$. }

\textbf{Proof.} Let $\varphi:C_m\otimes R_n\longrightarrow M_{mn}$,
$\varphi(c\otimes r)=cr$ be the isomorphism described
above. Then $\varphi$ maps bijectively the set of
decomposable tensors in $C_m\otimes R_n$ onto the
set of all rank 1 matrices. Consider $A'=\varphi^{-1}
A\varphi$. Then $A'$ is an automorphism of the
linear space $C_m\otimes R_n$, taking decomposable
tensors to decomposable ones. So $A'$ is a
decomposable automorphism by Proposition~10.

Any decomposable automorphism of $C_m\otimes R_n$ is
either of the form $c\otimes r\mapsto ac\otimes rb$, for
some $a\in GL_m$ and $b\in GL_n$; or $m=n$, and
the automorphism has the form $c\otimes r\mapsto
ar^t\otimes c^tb$, where $a,b\in GL_m$. In the first
case we have for matrices of the form $cr$
$$ A(cr)=(\varphi A'\varphi^{-1})(cr)=(\varphi A')
(c\otimes r)=\varphi(A'(c\otimes r))=\varphi
(ac\otimes rb)=acrb,$$
and therefore $Ax=axb$ for all $x$, because $x$
is a linear combination of rank 1 matrices.
In the second case
$$ A(cr)=\varphi(A'(c\otimes r))=\varphi(ar^t
\otimes c^tb)=ar^tc^tb=a(cr)^tb,$$
whence again $Ax=ax^tb$ for all~$x$.
\hfill $\square$ \medskip

Let $m,n,p\in{\Bbb N}$. Consider matrix multiplication
$M_{mn}\times M_{np}\longrightarrow M_{mp}$. In particular, for
any $x\in M_{mn}$ we can consider the subspace
$$ xM_{np}=\{xy\mid y\in M_{np}\}\subseteq M_{mp}.$$

\textbf{Lemma 12.} {\it ${\rm dim\,} xM_{np}=
p\cdot{\rm rk}(x)$ and ${\rm dim\,} M_{mn}
x'=m\cdot{\rm rk}(x')$ for all $x\in M_{mn}$ and
$x'\in M_{np}$. }

\textbf{Proof.} Prove the first equality; the second
can be considered similarly. Let $r={\rm rk}(x)$.
There exist nondegenerate matrices $a\in GL_m$
and $b\in GL_n$ such that $x=aE_rb$, where $E_r=
\sum_{i=1}^r e_{ii}$. Now $xM_{np}=aE_rb\cdot
M_{np}$. As the left multiplication by $a$ is an
invertible linear transfromation on $M_{mp}$, we
have
$$ {\rm dim\,} aE_rb\cdot M_{np}={\rm dim\,}
E_rb\cdot M_{np}. $$
Moreover, $bM_{np}=M_{np}$, so ${\rm dim\,}
xM_{np}={\rm dim\,} E_rM_{np}$. But the space
$E_rM_{np}$ is the
space of all $m\times p$ matrices that have zero
$j$-th rows for all $j\geq r+1$. Hence ${\rm dim\,} E_r
M_{np}=rp=p\cdot{\rm rk}(x)$.
\hfill $\square$ \medskip

\textbf{Proposition 13.} {\it Let $A$, $B$, and $C$
be linear
transformations of matrix spaces $M_{mn}$, $M_{np}$, and $M_{mp}$,
such that $A(x)B(y)=C(xy)$ for all $x\in M_{mn}$ and $y\in
M_{np}$. Then there exist $a\in GL_m$, $b\in GL_n$, and $c\in
GL_p$ such that $A(x)=axb$, $B(y)=b^{-1}yc$, $C(z)=azc$. }

\textbf{Proof.} For an arbitrary element $x\in M_{mn}$ and
a subspace $Y\subseteq M_{np}$ put
$$ xY=\{xy\mid y\in Y\}. $$
Also, for any two subspaces $X\subseteq M_{mn}$ and
$Y\subseteq M_{np}$ define
$$ XY=\langle xy\mid x\in X,\: y\in Y\rangle. $$
It is easy to deduce from the hypothesys that
always
$$ A(x)B(Y)=C(xY),\qquad A(X)B(Y)=C(XY). $$
In particular, let $x\in M_{mn}$ be an arbitrary
element, and $Y=M_{np}$. Then $A(x)B(Y)=A(x)M_{np}$,
and therefore $A(x)M_{np}=C(xM_{np})$.
Hence ${\rm dim\,} A(x)M_{np}={\rm dim\,} xM_{np}$,
whence $p\cdot{\rm rk}(A(x))=p\cdot{\rm rk}(x)$
by Lemma 12
and therefore ${\rm rk}(A(x))={\rm rk}(x)$. Thus, $A$
preserves the rank.

By Proposition 11, either there exist $a\in GL_m$
and $b\in GL_n$ such that $A(x)=axb$, or $m=n$ and $A(x)=ax^tb$, for all
$x\in M_{mn}$.

Admit the second possibility for $A$, and get a
contradiction. We may assume that $m=n\geq2$.
Take $X=e_1R_n$ and $Y=M_{np}$. Then $XY=e_1R_n
M_{np}=e_1R_p$, whence ${\rm dim\,} XY=p$. On the other
hand, $A(X)=a(e_1R_n)^tb=aC_ne^1b$, and further
$$A(X)B(Y)=A(X)M_{np}=(aC_ne^1b)(M_{np})=
(aC_ne^1)(bM_{np}).$$
As $b\in GL_n$, we have $bM_{np}=M_{np}$.
It is also clear that $e^1M_{np}=R_p$. So
$A(X)B(Y)=aC_nR_p=a\cdot M_{np}=M_{np}$;
in particular ${\rm dim\,} A(X)B(Y)=np$. So ${\rm dim\,} A(X)B(Y)
\ne{\rm dim\,} XY={\rm dim\,} C(XY)$, a contradiction.

Thus, $A(x)=axb$. The reader can prove similarly
that $B(y)=b_1yc_1$, for some $b_1\in GL_n$
and $c_1\in GL_p$.

Show that $b_1$ is proportional to $b^{-1}$. For
any $x\in M_{mn}$, $y\in M_{np}$ we have
$$ C(xy)=A(x)B(y)=axb\cdot b_1yc_1\,, $$
whence for any $d\in GL_n$
$$C(xy)=C(xd\cdot d^{-1}y)=axdb\cdot b_1d^{-1}yc_1.$$
So $axbb_1yc_1=axdbb_1d^{-1}yc_1$. As both $a$ and
$c_1$ are invertible, the latter equality implies
$xbb_1y=xdbb_1d^{-1}y$. Since this equality holds
for all $x\in M_{mn}$ and $y\in M_{np}$, it follows
that $bb_1=dbb_1d^{-1}$, for all $d\in GL_p$.
That is, $bb_1$ commutes with all elements of
$GL_p$ and so is a scalar matrix, $bb_1=\lambda
E_p$, $\lambda\in K^\ast$. That is, $b_1=\lambda
b^{-1}$.

Hence $B(y)=b_1yc_1=b^{-1}yc$, where $c=\lambda^{-1}
c_1$.

Thus, $A$ and $B$ can be defined by formulae
$A(x)=axb$, $B(x)=b^{-1}yc$. So $C(xy)=A(x)B(y)=
axyc$, for all $x\in M_{mn}$, $y\in M_{np}$.
As $M_{mn}M_{np}=M_{mp}$, we see that $C(z)=azc$,
for all $z\in M_{mp}$.
\hfill $\square$ \medskip

\textbf{6. Structure tensors and contragradient
maps.}
In this section we recall, briefly and without
proofs, some well-known concepts.

By $V^\ast$ we denote dual space of $V$, as usually.
As only finite-dimensional spaces are considered,
we identify $(V^\ast)^\ast$ with $V$.

For two elements $v\in V$ and $l\in V^\ast$ it
will be convenient to denote $l(v)$ either by
$\langle l,v\rangle$ or by $\langle v,l\rangle$. Thus, the element
$\langle u_1,u_2\rangle$ is defined, if one of the
elements $u_1$ and $u_2$ is in $V$, the other is
in $V^\ast$; and we always have $\langle u_1,u_2\rangle
=\langle u_2,u_1\rangle$. The symbol $\langle u_1,u_2\rangle$
is called the {\em pairing} of $u_1$ and $u_2$.

For any linear map $f:X\longrightarrow Y$ there exists a
unique linear map $f^\ast:Y^\ast\longrightarrow X^\ast$,
called the dual map, such that $\langle l,f(x)\rangle
=\langle f^\ast(l),x\rangle$ for all $x\in X$ and $l\in
Y^\ast$.

If $f$ is an isomorphism, then $f^\ast$ is an
isomorphism also, and $f^\vee=(f^\ast)^{-1}: X^\ast
\longrightarrow Y^\ast$ is called the map, {\em contragradient}
to~$f$. This is the unique map $X^\ast\longrightarrow Y^\ast$
satisfying the condition $\langle x,l\rangle=\langle f(x),
f^\ast(l)\rangle$ for all $x\in X$, $l\in X^\ast$.

If $f:X\longrightarrow Y$ and $g:Y\longrightarrow Z$ are linear maps,
then $(gf)^\ast=f^\ast g^\ast$. If $f$ and $g$
are isomorphisms, then $(gf)^\vee=g^\vee f^\vee$.
Also, $(f^\vee)^\vee=f$.

In particular, suppose that $\varphi:G\longrightarrow GL(X)$
is a representation of a group $G$ on a space~$X$.
Then the map $\varphi^\ast:G\longrightarrow GL(X^\ast)$,
defined by $\varphi^\ast(g)=\varphi(g)^\vee$, is a
representation also, called a representation
{\em contragradient} (or more often {\em dual})
to~$\varphi$.

Let $X$, $Y$, $Z$ be spaces. By ${\cal L}(X,Y)$ we
denote the space of all linear maps from $X$ to
$Y$, and by ${\cal L}_2(X,Y;Z)$ the space of all
bilinear maps $f:X\times Y\longrightarrow Z$. The spaces
${\cal L}(X,Y)$ and ${\cal L}_2(X,Y;Z)$ may be identified,
in a canonical way, with $X^\ast\otimes Y$ and
$X^\ast\otimes Y^\ast\otimes Z$, respectively (see
[11], \S 4.2). Describe this identification.
Let $l\in X^\ast$ and $y\in Y$. Consider the
map $\varphi_{l,y}:X\longrightarrow Y$, defined by
$$ \varphi_{l,y}(x)=l(x)y. $$
Clearly, $\varphi_{l,y}$ is a linear map. Furthermore,
the expression $l(x)y$ is linear in all three
arguments $l$, $x$, and $y$, and therefore the
rule $(l,y)\mapsto\varphi_{l,y}$ defines a bilinear
map from $X^\ast\times Y$ to ${\cal L}(X,Y)$. By the
universal property of tensor product there exists
a unique linear map $\varphi:X^\ast\otimes Y\longrightarrow{\cal L}(X,Y)$
such that $\varphi(l\otimes y)=\varphi_{l,y}$ for all $l$ and~$y$.
It can be shown (see [11]) that $\varphi$ is an
isomorphism.

We can define the isomorphism $\varphi:X^\ast\otimes
Y^\ast\otimes Z\longrightarrow{\cal L}_2(X,Y;Z)$ in a similar way.
Namely, $\varphi$ is the unique linear map such that
$$ (\varphi(l\otimes m\otimes z))(x,y)=l(x)m(y)z \quad \forall\
x\in X,\ y\in Y, z\in Z,\ l\in X^\ast,\ m\in
Y^\ast $$
(the details are left to the reader).

Let $f\in{\cal L}(X,Y)$ (resp., $f\in{\cal L}_2(X,Y;Z)$),
and let $h\in X^\ast\otimes Y$ (resp., $h\in X^\ast
\otimes Y^\ast\otimes Z$) be the tensor such that $\varphi(h)=f$.
This $h$ is called the {\em structure tensor}
of $f$, and will be denoted by~$\widetilde f$.

Consider the group $G=GL(X)\times GL(Y)$. It acts
on the spaces $X^\ast\otimes Y$ and ${\cal L}(X,Y)$ as
usually. That is, an element $g=(g_1,g_2)\in G$
acts on $X^\ast\otimes Y$ as $g_1^\vee\otimes g_2$, and
the action of $g$ on ${\cal L}(X,Y)$ is defined by $g(f)=
g_2fg_1^{-1}$ (we leave to the reader to show that
this is indeed a left action). Similarly, the
group $G=GL(X)\times GL(Y)\times GL(Z)$ acts on
$X^\ast\otimes Y^\ast\otimes Z$ and on ${\cal L}_2(X,Y;Z)$.
The element $g=(g_1,g_2,g_3)\in G$ acts on
$X^\ast\otimes Y^\ast\otimes Z$ as $g_1^\vee\otimes g_2^\vee
\otimes g_3$, and the action on ${\cal L}_2(X,Y;Z)$ is
described by the rule
$$ (g(f))(x,y)=g_3(f(g_1^{-1}x, g_2^{-1}y)) $$
(i.e., $g$ takes $f$ to the map $f_1$ defined
by $f_1(x,y)=g_3(f(g_1^{-1}x,g_2^{-1}y))$; we
may also write this as $g(f)=g_3\circ f\circ
(g_1^{-1}\times g_2^{-1})$).

The following proposition is well known.

\textbf{Proposition 14.} {\it
Let $G=GL(X)\times GL(Y)$ (resp. $G=GL(X)\times
GL(Y)\times GL(Z)$), and let $\varphi:X^\ast\otimes Y
\longrightarrow{\cal L}(X,Y)$ (resp. $\varphi:X^\ast\otimes Y^\ast\otimes Z
\longrightarrow{\cal L}_2(X,Y;Z)$) be the canonical isomorphism.
Then $\varphi$ is an isomorphism of $KG$-modules. }

\textbf{7. The isotropy group of a bilinear
map.}
Let $X$, $Y$, and $Z$ be vector spaces and let
$f\in{\cal L}_2(X,Y;Z)$ be a bilinear map. The group
$G=GL(X)\times GL(Y)\times GL(Z)$ acts on ${\cal L}_2
(X,Y;Z)$ in the way described in the end of the
previous section. The stabilizer of $f$ in $G$ with
respect to this action will be called the {\em
isotropy group} of $f$, and will be denoted
by~$\Delta(f)$. The reader can easily check that
this definition is equivalent to the following:
$\Delta(f)$ is the set of all triples $(A,B,C)\in
G$ such that $f(Ax,By)=Cf(x,y)$ for all $x\in X$
and $y\in Y$. In other words, the diagram
$$ \begin{CD}  X\times Y @>f>> Z \\ @V{A\times B}VV
@ VV{C}V \\ X\times Y @>f>> Z \end{CD} $$
must commute.

\textbf{Example.} Let $U$, $V$, and $W$ be three
spaces, let $X={\cal L}(U,V)$, $Y={\cal L}(V,W)$,
$Z={\cal L}(U,W)$,
and let $f: X\times Y\to Z$ be the usual
composition of mappings, i.e., $f(x,y)=yx$.
Clearly $f$ is bilinear. For $g=(g_1,g_2,g_3)\in
GL(U)\times GL(V)\times GL(W)$ put $R(g)=(A,B,C)$,
where $A\in GL(X)$, $B\in GL(Y)$ and $C\in GL(Z)$
are defined by $Ax=g_2xg_1^{-1}$,  $By=g_3y
g_2^{-1}$, and $Cz=g_3zg_1^{-1}$, respectively.
Then it is easy to see that $R(g)\in\Delta(f)$
for all~$g$, and Proposition 13 actually shows that
$\Delta(f)=\{ R(g)\mid g\in G\}$.

The following proposition shows that the isotropy
group of a bilinear mapping is closely related to
the (small) isotropy group of the corresponding
structure tensor.

\textbf{Proposition 15.} {\it
Let $f:X\times Y\to Z$ be a bilinear mapping
and let $\widetilde f\in X^\ast\otimes Y^\ast
\otimes Z$ be its
structure tensor. Let $(A,B,C)\in G=GL(X)\times
GL(Y)\times GL(Z)$. Then $(A,B,C)\in\Delta(f)$
if and only if $A^\vee\otimes B^\vee\otimes C\in
\Gamma^0(\widetilde f)$. }

\textbf{Proof.} By Proposition 14, the map
$h\mapsto
\widetilde h$ is a $G$-isomorphism from ${\cal L}_2(X,Y;Z)$
to $X^\ast\otimes Y^\ast\otimes Z$. So $g=(A,B,C)\in G$
is in $\Delta(f)$ if and only if $g$ fixes~$
\widetilde f$. But $g(\widetilde f)=(A^\vee\otimes B^\vee\otimes C)
\widetilde f$ by the definition of the action of $G$
on $X^\ast\otimes Y^\ast\otimes Z$.
\hfill $\square$ \medskip

\textbf{8. Proof of Proposition 5.}
We start with the following observation. Let $x$
and $y$ be $a\times b$ and $b\times a$ matrices,
respectively. Then ${\rm Tr}(xy)={\rm Tr}(yx)$. Moreover,
$$ \langle x,y\rangle ={\rm Tr}(xy)={\rm Tr}(yx) $$
is a nondegenerate bilinear pairing between
$M_{ab}$ and~$M_{ba}$. Therefore we may identify
$M^\ast_{ab}$ with $M_{ba}$, and $M^\ast_{ba}$
with~$M_{ab}$.

Further, the group $G=GL_a\times GL_b$ acts
on both $M_{ab}$ and $M_{ba}$ in a usual way,
that is, $g=(g_1,g_2)$ takes $x\in M_{ab}$ and
$y\in M_{ba}$ to $g_1xg_2^{-1}$ and $g_2yg_1^{-1}$,
respectively. The pairing is invariant under this
action. Indeed, if $x\in M_{ab}$, $y\in M_{ba}$,
and $g=(g_1,g_2)\in G$, then
$$\langle gx,gy\rangle={\rm Tr}((g_1xg_2^{-1})
(g_2yg_1^{-1}))={\rm Tr}(g_1xyg_1^{-1})={\rm Tr}
(xy)=\langle x,y\rangle.$$
Therefore the transformations, induced by $g$
on $M_{ab}$ and $M_{ba}$, are contragradient
each to the other.

Let $L_1=M_{mn}$, $L_2=M_{np}$ and $L_3=M_{pm}$
be as in the hypothesis of the Proposition, and
let $N_1=M_{nm}$ and $N_2=M_{pn}$. Then $N_i$ is
dual to $L_i$, $i=1,2$. Let $\varphi:N_1\times N_2
\longrightarrow L_3$ be the usual product map, that is,
$\varphi(x,y)=yx$. Its structure tensor $\widetilde\varphi\in
N_1^\ast\otimes N_2^\ast\otimes L_3$ may be considered
as an element of $L_1\otimes L_2\otimes L_3$. We show
that $\widetilde\varphi=t=\langle m,n,p\rangle$.

Indeed, we have
$$ t=\sum_{(i,j,k)\in\overline m\times\overline n
\times\overline p} e_{ij}\otimes e_{jk} \otimes
e_{ki}\,. $$
Let
$$\psi:L_1\otimes L_2\otimes L_3=N^\ast_1\otimes
N^\ast_2\otimes L_3 \longrightarrow {\cal L}_2
(N_1,N_2;L_3) $$
be the canonical map, described in Section~6
(denoted by $\varphi$ there). We must show that the
bilinear map $\rho=\psi(t)$ coincides with~$\varphi$.
The bases of $N_1$ and $N_2$ are $\{e_{uv}\mid
u\in\overline n,\ v\in\overline m\}$ and
$\{e_{wq} \mid w\in\overline p,\ q\in\overline n\}$,
respectively. It follows from the definition of
$\psi$ that the value of $\rho$ on the pair
$(e_{uv},e_{wq})$ equals
\begin{eqnarray*}
\sum_{i,j,k} \langle e_{ij},e_{uv}\rangle \langle
e_{jk},e_{wq}\rangle e_{ki} &=& \sum_{i,j,k}
{\rm Tr}(e_{ij}e_{uv}) {\rm Tr}(e_{jk}e_{wq})\,
e_{ki} \\
&=& \sum_{i,j,k} \delta_{iv}\delta_{ju}\delta_{kw}
\delta_{jq} e_{ki}=\delta_{uq}e_{wv}\,,
\end{eqnarray*}
where the sum is taken over all $(i,j,k)\in
\overline m\times\overline n\times\overline p$.

On the other hand, $\varphi(e_{uv},e_{wq})=e_{wq}
e_{uv}=\delta_{uq}e_{wv}$. Thus, $\rho(e_{uv},
e_{wq})=\varphi(e_{uv},e_{wq})$ for all $u$, $v$,
$w$, $q$, that is, $\varphi=\rho$. Thus,
$t=\widetilde\varphi$.

Return to the proof of the Proposition, and assume
that $A\in\Gamma^0(t)$. We have $A=A_1\otimes A_2
\otimes A_3$, for some $A_i\in GL(L_i)$, $i=1,2,3$.
For $i=1,2$ we put $B_i=A_i^\vee\in GL(L_i^\ast)
=GL(N_i)$. Then $A_i=B_i^\vee$, $i=1,2$. So we
have $B_1^\vee\otimes B_2^\vee\otimes A_3\in\Gamma^0(\widetilde
\varphi)$. Now Proposition 15 implies that $(B_1,
B_2,A_3)\in\Delta(\varphi)$. In other words, $(B_2y)
(B_1x)=A_3(yx)$ for any $x\in M_{nm}$ and $y
\in M_{pn}$.

By Proposition 13, there exist $a_1\in GL_p$,
$b_1\in GL_n$, and $c_1\in GL_m$ such that
$B_2y =a_1yb_1$, $B_1x=b_1^{-1}xc_1$, and
$A_3z=a_1zc_1$, for all $x\in N_1$, $y\in N_2$,
$z\in L_3$.

It follows from the discussion in the beginning
of the proof that the transformation on $L_1$,
contragradient to the transformation $x\mapsto
b_1^{-1}xc_1$ on $N_1$, may be described by the
formula $x'\mapsto c_1^{-1}x'b_1$. Similarly $A_2$
acts by the rule $y'\mapsto b_1^{-1}y'a_1^{-1}$.
Therefore $A$ acts by
$$ A(x\otimes y\otimes z)=c_1^{-1}xb_1\otimes
b_1^{-1}ya_1^{-1}\otimes a_1zc_1\,. $$
That is,
$$ A(x\otimes y\otimes z)=axb^{-1}\otimes byc^{-1}
\otimes cza^{-1}, $$
where $a=c_1^{-1}$, $b=b_1^{-1}$, and $c=a_1$.
Thus, $A=T(a,b,c)$.
\hfill $\square$ \medskip

\bigskip
\centerline{References}
\medskip

1. Burichenko V.P., On symmetries of the Strassen
algorithm // arXiv: 1408.6273, 2014. 

2. Burichenko V.P., Symmetries of matrix multiplication
algorithms. I / arXiv: 1508.01110, 2015.

3. Strassen V., Gaussian elimination is not
optimal. Numer.Math. 13 (1969), 354--356.

4. Hopcroft J.E., Kerr L.R., On minimizing the number of multiplications
necessary for matrix multiplication. SIAM J.Appl.Math. 20(1971), 
30--36. 

5. Laderman J., A noncommutative algorithm for
multiplying $3\times3$ matrices using 23 multiplications.
Bull.Amer.Math.Soc. 82(1976), 126--128. 

6. Burichenko V.P., Symmetries of matrix multiplication algorithms // (In
preparation) 

7. de Groote H.F., On the varieties of optimal algorithms for the computation 
of bilinear mappings. I. The isotropy group of a bilinear mapping. // 
Theor.Comput.Sci. 7 (1978), 1--24. 

8. de Groote H.F., On the varieties of optimal algorithms for the 
computation of bilinear mappings. II. Optimal algorithms for $2\times2$ 
matrix multiplication // Theor.Comput.Sci. 7 (1978), 127--148. 

9. Brockett R.W., Dobkin D., On the optimal evaluation of a set of bilinear 
forms // Linear Algebra Appl. 19(1978), 207--235. 

10. Chiantini L., Ikenmeyer C., Landsberg J.M., Ottaviani G., The geometry of rank
decompositions of matrix multiplication I: $2\times2$ matrices //
 arXiv: 1610.08364v1. 2016. 

11. Kostrikin A.I., Manin Yu.I., Linear Algebra and Geometry.
2nd ed.  Gordon and Breach, 1997. 

12. Kostrikin A.I., Introduction to Algebra. Moscow: Nauka, 1977 (in Russian).

13. Curtis C.W., Reiner I., Representation Theory of Finite Groups and Associative 
Algebras. Interscience Publishers, 1962.

14. Burau W., Mehrdimensionale Projective und Hohere Geometrie: Deutsche Verlag 
der Wissenschaften, Berlin, 1961.

15. Grochow J.A., Moore C., Matrix multiplication algorithms from group orbits / 
arXiv 1612.01527v1. 2016. 

\end{document}